\documentclass[useAMS,usenatbib]{mn2e}

\usepackage{natbib}
\usepackage{graphicx}
\usepackage{amssymb}
\usepackage{color}
\usepackage{caption}
\usepackage{lipsum,graphicx,multicol}
\usepackage{float}

\usepackage{subcaption}

\title[Radiative dusty feedback in early quasars]{AGN radiative feedback in the early growth of massive black holes}

\author[ ]
{W. Ishibashi$^{1}$\thanks{E-mail: wako.ishibashi@physik.uzh.ch} 
\vspace{0.05cm}
\footnotemark[0]\\
$^{1}$Physik-Institut, Universitat Zurich, Winterthurerstrasse 190, 8057 Zurich, Switzerland 
}

\begin{document}

\pdfminorversion=4

\date{Accepted ? Received ?; in original form ? }

\pagerange{\pageref{firstpage}--\pageref{lastpage}} \pubyear{2012}

\maketitle

\label{firstpage}

\begin{abstract} 
Growing observational evidence confirms the existence of massive black holes ($M_{BH} \sim 10^9 M_{\odot}$), accreting at rates close to the Eddington limit, at very high redshifts ($z \gtrsim 6-7$) in the early Universe. Recent observations indicate that the host galaxies of the first quasars are chemically evolved systems, containing unexpectedly large amounts of dust. Such a combination of high luminosities and large dust content should form favourable physical conditions for radiative dusty feedback. We explore the impact of the active galactic nucleus (AGN) feedback, driven by radiation pressure on dust, on the early growth of massive black holes. Assuming Eddington-limited exponential black hole growth, we find that the dynamics and energetics of the radiation pressure-driven outflows also follow exponential trends at late times. We obtain modest outflow energetics (with momentum flux $\dot{p} \lesssim L/c$ and kinetic power $\dot{E}_{k} \lesssim 10^{-3} L$), comparable with available observations of quasar-driven outflows at very high redshifts, but significantly lower than typically observed in local quasars and predicted by wind energy-driven models. AGN radiative dusty feedback may thus play an important role in powering galactic outflows in the first quasars in the early Universe. 
\end{abstract} 



\section{Introduction}

Observations over the past years have revealed the existence of massive black holes, with $M_{BH} \sim 10^9 M_{\odot}$, at very high redshifts of $z \sim 6-7$ \citep{Mortlock_et_2011, Wu_et_2015, Banados_et_2018}. 
To date, more than hundred quasars are known at $z > 6$ \citep[][]{Gallerani_et_2017, Wang_et_2018, Yang_et_2018}. The recent discovery of ULAS J1342+0928 sets the current distance record at a redshift of $z = 7.5$ \citep{Banados_et_2018}. The very existence of such massive black holes in the early Universe, when its age was less than $\lesssim 1$ Gyr, poses severe challenges to black hole formation models.  

A number of scenarios have been proposed for the formation of black hole seeds and their early growth. Three main pathways have been discussed in the literature: as the remnants of the first generation of Population III stars with primordial gas composition ($M_{BH,0} \sim 10^2-10^3 M_{\odot}$), as the result of runaway collisions in nuclear star clusters ($M_{BH,0} \sim 10^3-10^4 M_{\odot}$), and as the outcome of direct collapse of gas without fragmentation or DCBH ($M_{BH,0} \sim 10^4-10^6 M_{\odot}$) \citep[][and references therein]{Volonteri_2010, Valiante_et_2017, Smith_et_2017, Wise_2018}. In the case of light seeds, a continuous growth at the Eddington rate and/or several super-Eddington accretion episodes are required, while the constraints may be slightly relaxed in the case of heavy seeds. In any case, a rapid and efficient black hole growth is required at early times in order to account for the billion-solar mass objects observed at $z \gtrsim 6$. 

The highest-redshift quasars are characterised by high bolometric luminosities ($L \gtrsim 10^{47}$erg/s) and high accretion rates, possibly accreting close to the Eddington limit ($L/L_E \sim 1$) \citep{Banados_et_2018, Mazzucchelli_et_2017}. Recent observations have also revealed, somewhat unexpectedly, that the host galaxies of the first quasars are chemically evolved systems containing metal-enriched gas with super-solar metallicity \citep[][and references therein]{Calura_et_2014, Valiante_et_2017}. Indeed, a significant amount of dust, of the order of $M_{dust} \gtrsim 10^8 M_{\odot}$, is detected in the most distant quasar at $z = 7.5$ \citep{Venemans_et_2017}.  
ALMA observations of a sample of quasars at $z \gtrsim 6$ also indicate dust masses in the range $M_{dust} \sim (10^7-10^9) M_{\odot}$ \citep{Venemans_et_2018}. This suggests that considerable quantities of dust are already present in the quasar host galaxies at very high redshifts, and thus early enrichment must be common. 

Combining the above two observational facts, i.e. the high luminosity output of the first quasars and the presence of large amounts of dust in their host galaxies, we expect AGN radiative feedback, due to radiation pressure on dust, to play a major role in the early black hole growth phase. We have previously discussed how AGN radiative feedback may drive powerful outflows on galactic scales, and how the observed outflow dynamics and energetics may be accounted for, by including radiation trapping \citep{Ishibashi_Fabian_2015, Ishibashi_et_2018}. 
More recently, we have also analysed the effects of AGN luminosity decay on the inferred outflow energetics \citep{Ishibashi_Fabian_2018}. Here we investigate for the first time the impact of AGN radiative feedback on the early growth of massive black holes, by considering a physically motivated luminosity output (powered by an exponentially growing black hole) and analyse the resulting outflow dynamics and energetics. 


\section{Radiative dusty feedback driven by Eddington-limited growth}

We consider AGN radiative feedback driven by radiation pressure on dust \citep[cf.][]{Ishibashi_Fabian_2015, Ishibashi_et_2018}. Radiation pressure sweeps up the surrounding dusty gas into an outflowing shell. 
The equation of motion of the shell is given by:
\begin{equation}
\frac{d}{dt} [M_{g}(r) v] = \frac{L(t)}{c} (1 + \tau_{IR} - e^{-\tau_{UV}} ) - \frac{G M(r) M_{g}(r)}{r^2}
\end{equation} 
where $L(t)$ is the central luminosity evolution, $M(r)$ is the total mass distribution, $M_g(r)$ is the gas mass, and $\tau_\mathrm{IR,UV}$ are the infrared (IR) and ultraviolet (UV) optical depths.  

Assuming Eddington-limited accretion, the black hole mass grows exponentially as $M_\mathrm{BH}(t) \cong M_0 e^{t/t_\mathrm{S}}$, where $M_0$ is the initial mass and $t_\mathrm{S} = \frac{\epsilon \sigma_T c}{4 \pi G m_p} \sim 5 \times 10^7$yr is the Salpeter time. As a consequence, the luminosity output also grows exponentially as 
\begin{equation}
L(t) \cong L_0 e^{t/t_\mathrm{S}}
\end{equation}
where $L_0$ is the initial luminosity. 
We assume the simple case of an isothermal potential:
\begin{equation}
M(r) = \frac{2 \sigma^2 r}{G} 
\end{equation} 
where $\sigma$ is the velocity dispersion. 
The ambient gas density distribution can be parameterised as a power-law of radius: 
$n(r) = n_0 (\frac{r}{R_0})^{-\alpha}$, where $\alpha$ is the power-law exponent, $n_0$ is density of the external medium, and $R_0$ is the initial radius. The corresponding shell mass is given by: $M_g(r) = 4 \pi m_p n_0 R_0^{\alpha} \frac{r^{3-\alpha}}{3-\alpha}$. 
We consider here an isothermal gas distribution, corresponding to the case $\alpha = 2$: 
\begin{equation}
M_g(r) = 4 \pi m_p n_0 R_0^2 r
\label{Eq_gas_mass}
\end{equation}
The associated IR and UV optical depths are given by: 
\begin{equation}
\tau_\mathrm{IR}(r) = \frac{\kappa_\mathrm{IR} M_g(r)}{4 \pi r^2} = \frac{\kappa_\mathrm{IR} m_p n_0 R_0^2}{r}
\end{equation}
\begin{equation}
\tau_\mathrm{UV}(r) = \frac{\kappa_\mathrm{UV} M_g(r)}{4 \pi r^2} = \frac{\kappa_\mathrm{UV} m_p n_0 R_0^2}{r}
\end{equation} 
where $\kappa_\mathrm{IR}$ and $\kappa_\mathrm{UV}$ are the IR and UV opacities, respectively. 
In the following, we take as fiducial parameters: $L_0 = 10^{44}$ erg/s, $n_0 = 10^4 cm^{-3}$, $\sigma = 50$ km/s, $R_0 = 3$ pc, $\kappa_\mathrm{IR} = 5 \, \mathrm{cm^2 g^{-1} f_{dg, MW}}$, $\kappa_\mathrm{UV} = 10^3 \, \mathrm{cm^2 g^{-1} f_{dg, MW}}$, with the dust-to-gas ratio normalised to the Milky Way value. 
We note that the fiducial initial luminosity would correspond to an initial black hole mass of $\sim 10^6 M_{\odot}$ radiating at the Eddington limit.


\subsection{Radiative and gravitational forces}

The evolution of the shell is governed by the balance between the radiative and gravitational forces. 
The outward force due to radiation pressure is given by: 
\begin{equation}
F_\mathrm{rad} = \frac{L(t)}{c} (1 + \tau_\mathrm{IR} - e^{-\tau_\mathrm{UV}} ) 
\end{equation}
\begin{equation}
= \frac{L_0 \exp(t/t_\mathrm{S})}{c} (1 + \frac{\kappa_\mathrm{IR} m_p n_0 R_0^2}{r} - \exp(-\frac{\kappa_\mathrm{UV} m_p n_0 R_0^2}{r}))
\end{equation} 
The inward force due to gravity is given by:
\begin{equation}
F_\mathrm{grav} = \frac{G M(r) M_{g}(r)}{r^2} = 8 \pi m_p n_0 R_0^2 \sigma^2 
\end{equation} 
which is a constant, independent of radius.

\begin{figure}
\begin{center}
\includegraphics[angle=0,width=0.4\textwidth]{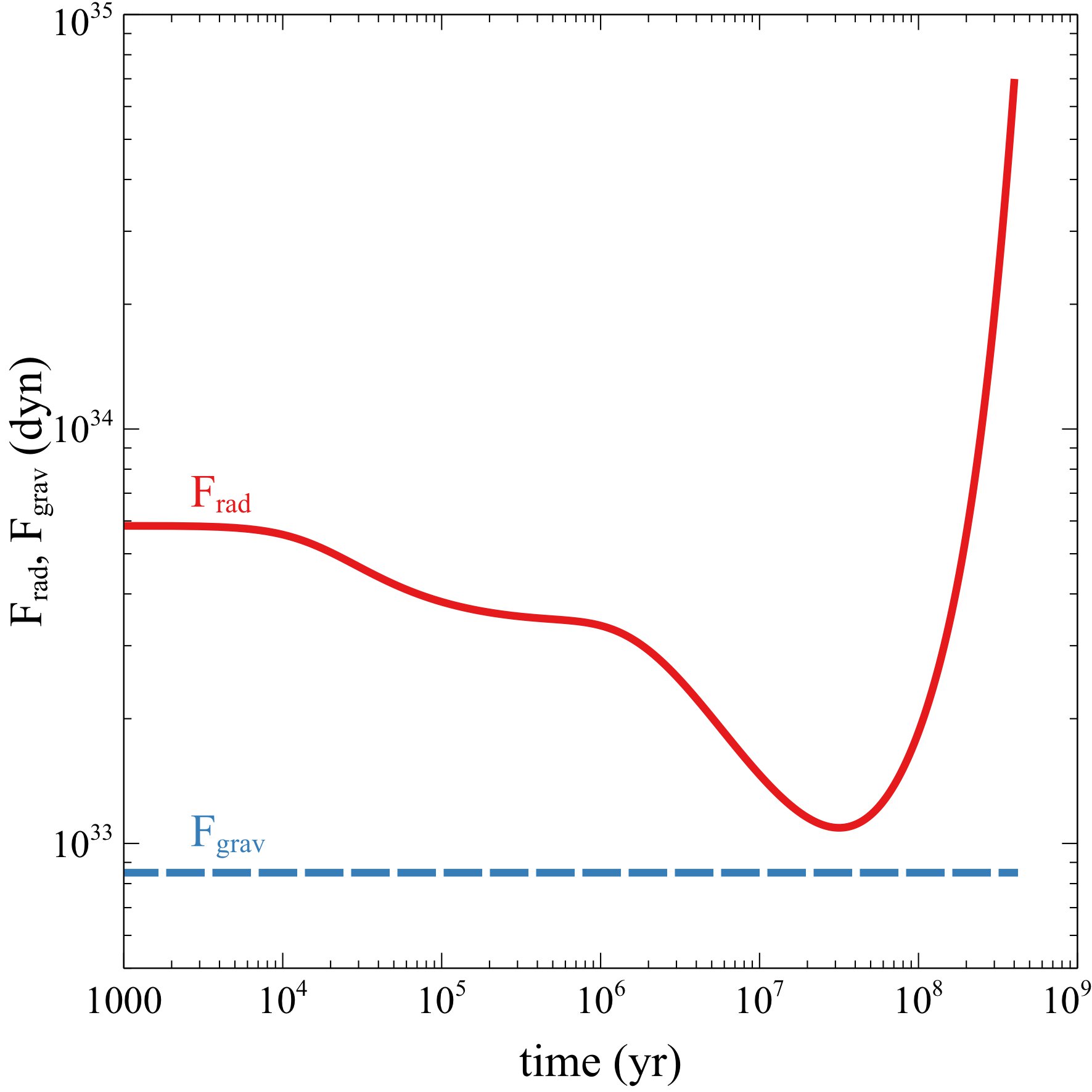} 
\caption{ 
Radiative force (red solid) and gravitational force (blue dashed) as a function of time. 
}
\label{Fig_FradFgrav_t}
\end{center}
\end{figure}

In Fig. $\ref{Fig_FradFgrav_t}$, we plot the radiative and gravitational forces as a function of time.
At early times ($t \ll t_\mathrm{S}$), the luminosity can be approximated as a constant $L(t) \cong L_0$; while at late times ($t \gtrsim t_\mathrm{S}$), the exponential term takes over and the luminosity grows as $L(t) \propto \exp(t/t_\mathrm{S})$. 
As a result, the radiative force reaches a minimum around the Salpeter time ($t \lesssim t_\mathrm{S}$), and then increases exponentially. 

\begin{figure}
\begin{center}
\includegraphics[angle=0,width=0.4\textwidth]{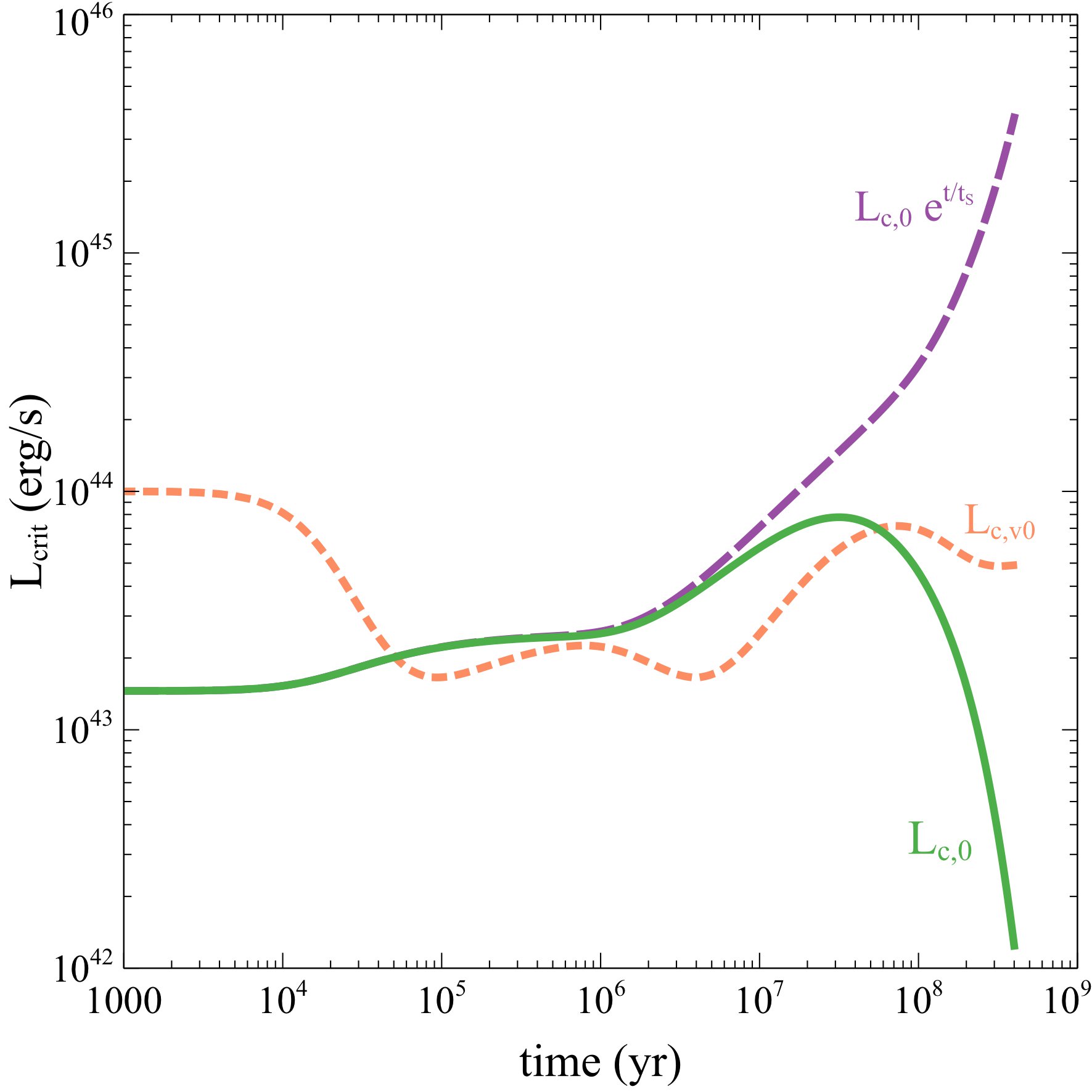} 
\caption{ 
Critical luminosities as a function of time: $L_{c,0}$ (green solid), $L_{c,0} \exp(t/t_S)$ (violet dashed), $L_{c,v0}$ (orange dotted). 
}
\label{Fig_Lc0_t}
\end{center}
\end{figure}

A critical luminosity can be defined by equating the outward force due to radiation pressure to the inward force due to gravity ($F_{rad} = F_{grav}$):
\begin{equation}
L_{c,0} = \frac{1}{\exp(t/t_\mathrm{S})} \frac{8 \pi c m_p n_0 R_0^2 \sigma^2}{(1 + \tau_\mathrm{IR} - e^{-\tau_\mathrm{UV}} ) } \, , 
\end{equation} 
with the corresponding temporal evolution $L_{c,0} \exp(t/t_\mathrm{S})$. 
Alternatively, one may also define another critical luminosity based on the condition that the shell velocity crosses zero (v=0): $L_{c,v0}$. 
In Fig. $\ref{Fig_Lc0_t}$, we show the different critical luminosities as a function of time.  
We see that the critical luminosity $L_{c,0}$ reaches a maximum around $t \lesssim t_\mathrm{S}$ (green solid line), corresponding to the minimum in $F_\mathrm{rad}$ (cf. Fig. $\ref{Fig_FradFgrav_t}$). Provided that this minimum radiative force exceeds the gravitational force ($F_\mathrm{rad,min} > F_\mathrm{grav}$), a steady outflow can develop on large scales. 

\section{Outflow dynamics and energetics}

We next consider the propagation of radiation pressure-driven outflows powered by exponentially growing black holes, and the resulting outflow dynamics and energetics. 


\subsection{Outflow dynamics}

\begin{figure}
\begin{center}
\includegraphics[angle=0,width=0.4\textwidth]{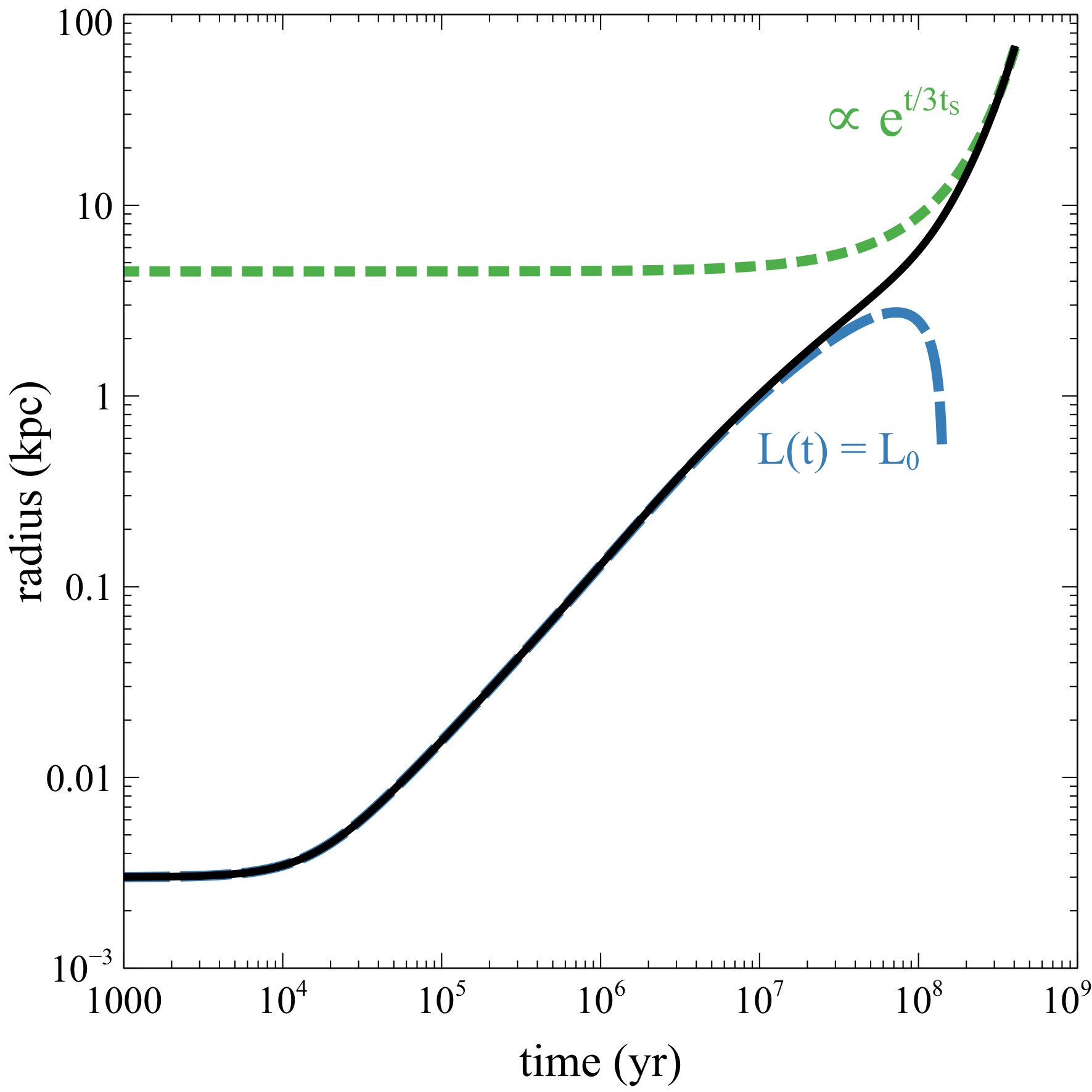} 
\caption{Shell radius (black solid) as a function of time, radius evolution for a constant luminosity output (blue dashed), and analytic exponential trend (green dotted). 
}
\label{Fig_r_t}
\end{center}
\end{figure}

In Fig. $\ref{Fig_r_t}$, we plot the evolution of the shell radius as a function of time. 
At early times ($t \ll t_\mathrm{S}$), the temporal evolution of the shell radius is equivalent to the case of a constant luminosity output ($L(t) = L_0$). At late times ($t > t_\mathrm{S}$), the exponential term takes over, and the shell evolution becomes purely exponential: $r(t) \propto \exp(t/3t_\mathrm{S})$. 
We note that the shell expansion has an e-folding time of 3 times the Salpeter time, and is thus slower than the luminosity output ($L(t) \propto \exp(t/t_\mathrm{S})$). 

\begin{figure}
\begin{center}
\includegraphics[angle=0,width=0.4\textwidth]{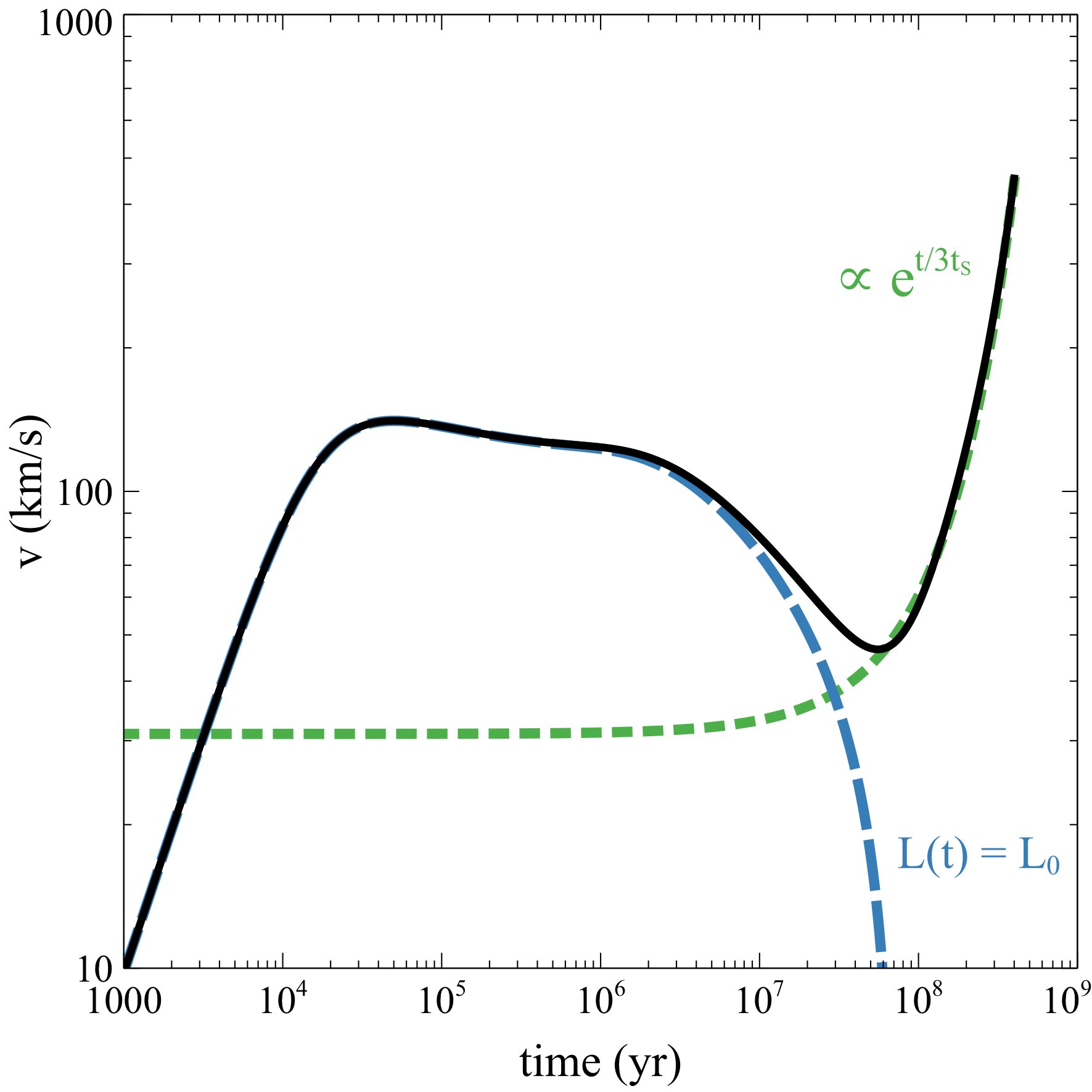} 
\caption{Shell velocity (black solid) as a function of time, velocity evolution for a constant luminosity output (blue dashed), and analytic exponential trend (green dotted). 
}
\label{Fig_v_t}
\end{center}
\end{figure}

The corresponding shell velocity as a function of time is shown in Fig. $\ref{Fig_v_t}$. 
At early times, the velocity evolution is the same as in the case of a constant luminosity output. 
The shell speed reaches a minimum around the Salpeter time ($t \gtrsim t_\mathrm{S}$), and then increases exponentially as $v(t) \propto \exp(t/3t_\mathrm{S})$.  
We note that the shell velocity has the same exponential dependence as the shell radius, corresponding to a linear increase in radius: $v \propto r$. 

\subsection{Outflow energetics}

\begin{figure*}
\begin{multicols}{3}
    \includegraphics[width=\linewidth]{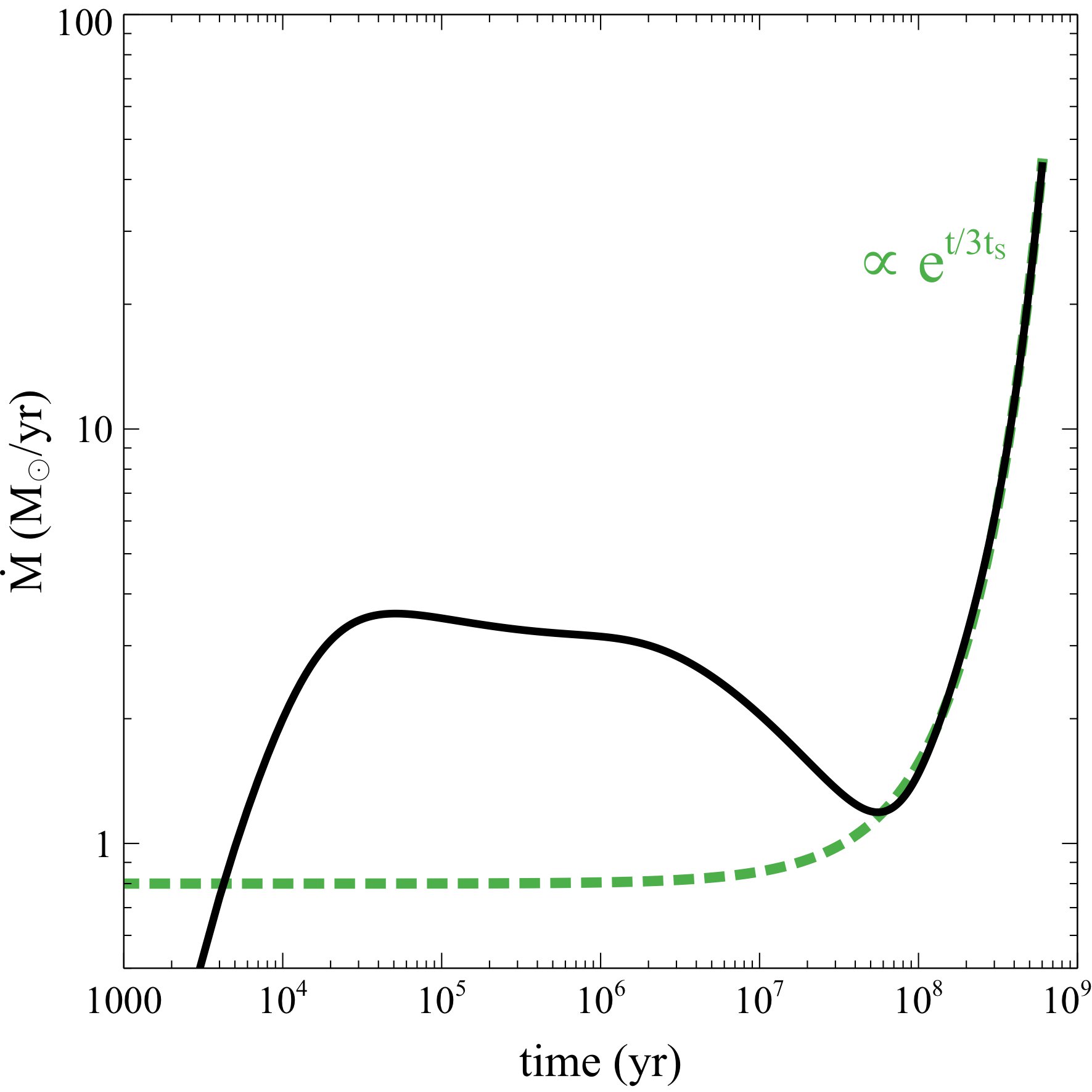}\par 
    \includegraphics[width=\linewidth]{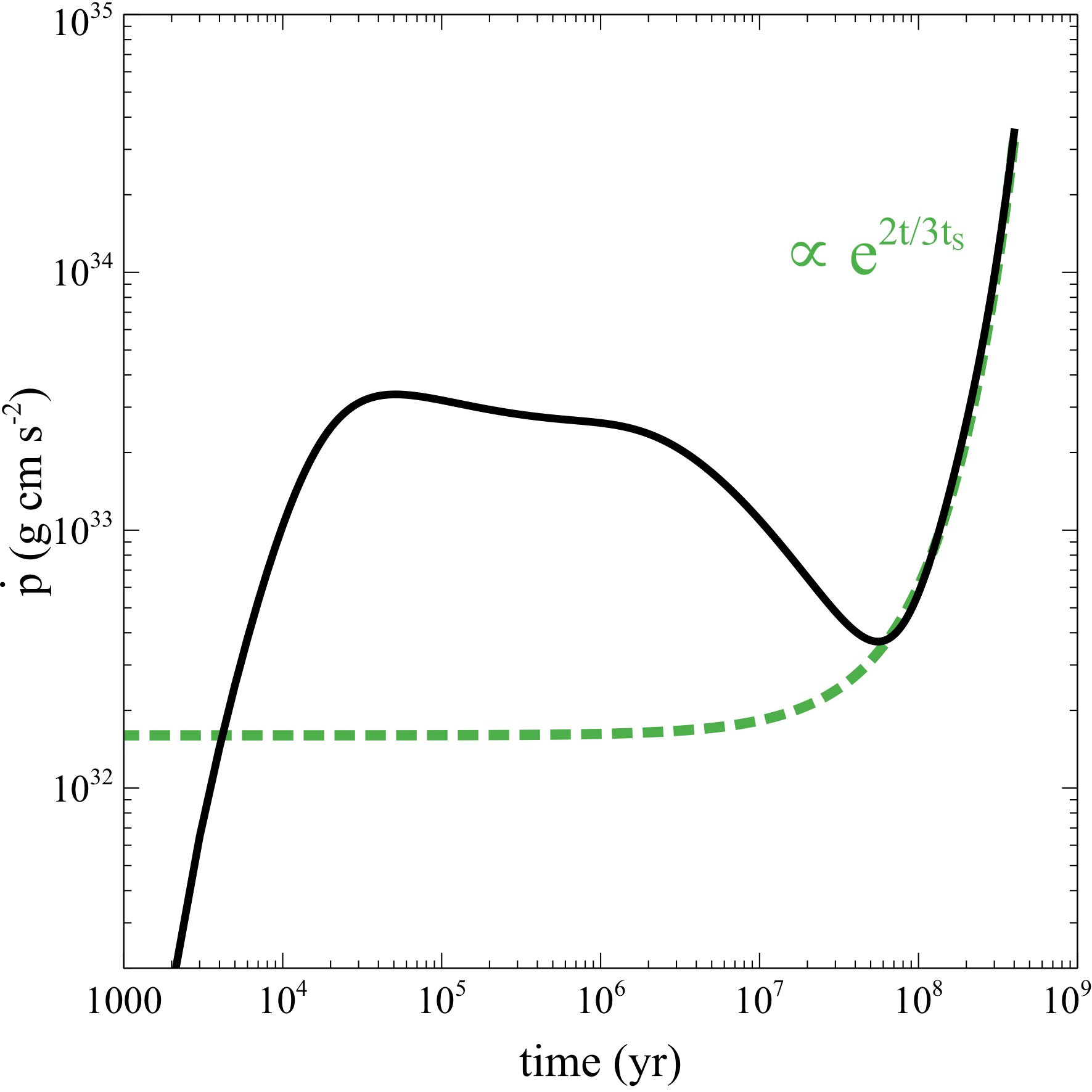}\par 
    \includegraphics[width=\linewidth]{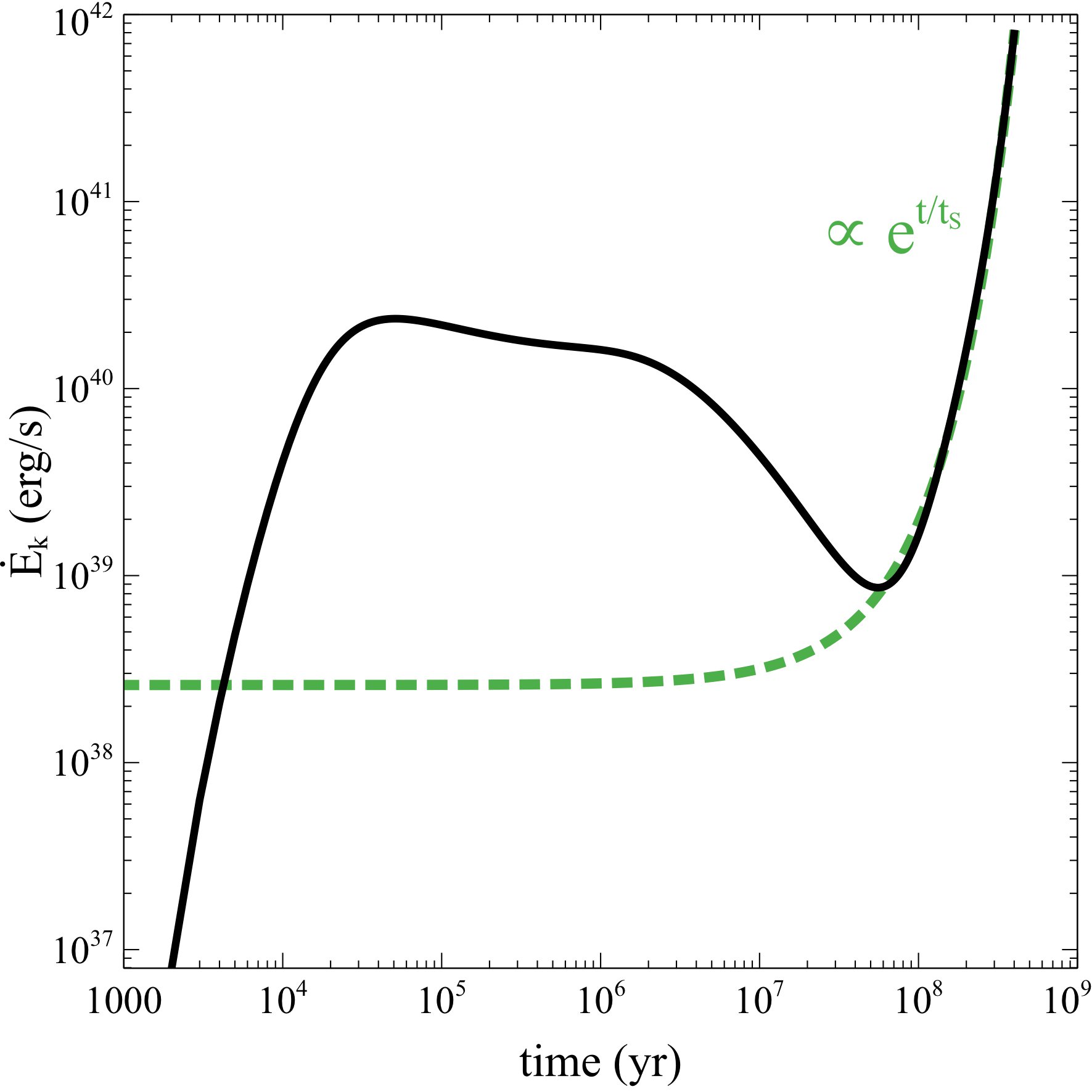}\par
    \end{multicols}
\caption{ Outflow energetics as a function of time (black solid): mass outflow rate $\dot{M}$ (left panel), momentum flux $\dot{p}$ (middle panel), and kinetic power $\dot{E}_{k}$ (right panel), with analytic exponential trends (green dotted). 
}
\label{Fig_MpEkdot_t}
\end{figure*}

\begin{figure}
\begin{center}
\includegraphics[angle=0,width=0.4\textwidth]{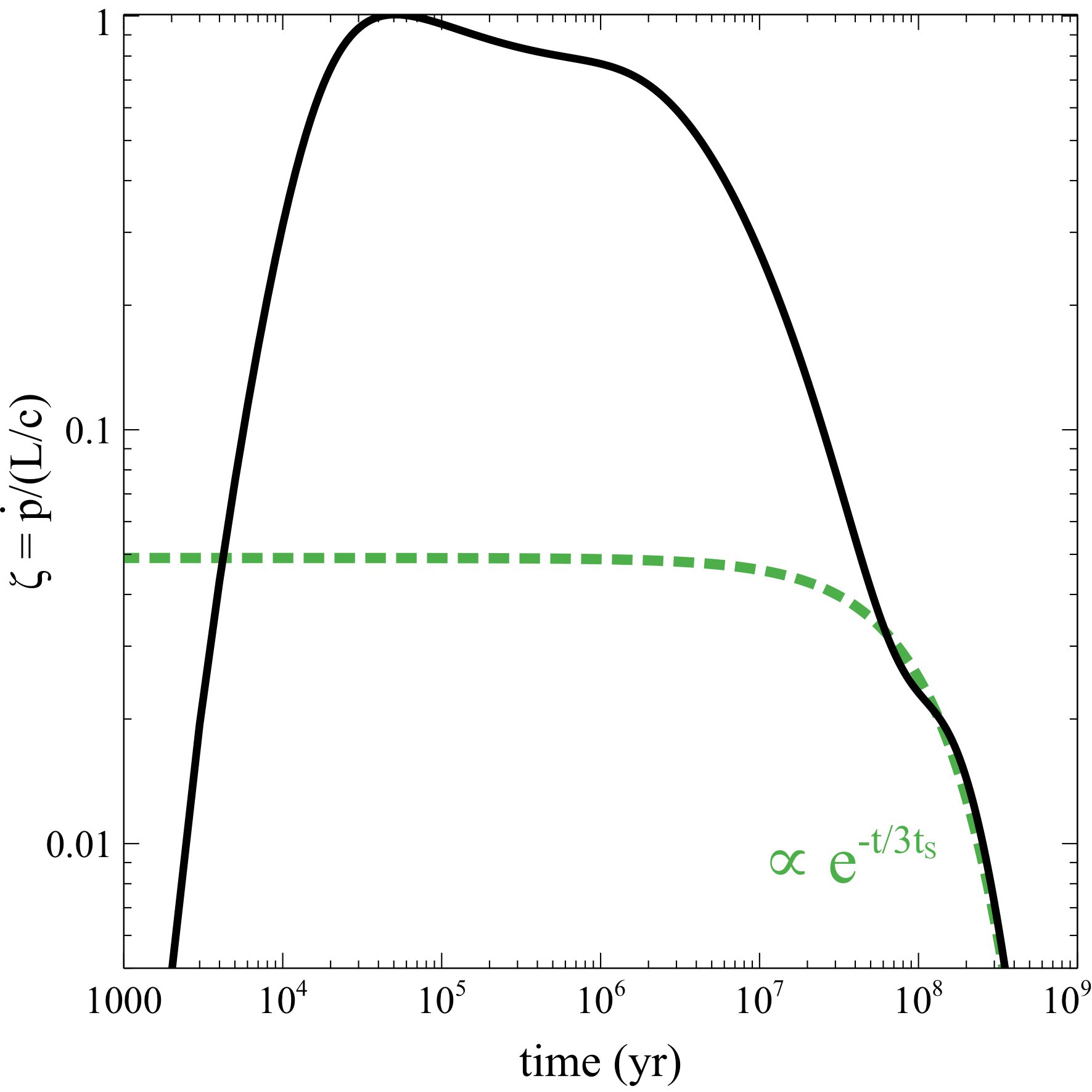} 
\caption{
Momentum ratio versus time (black solid), with analytic exponential trend (green dotted). 
 }
\label{Fig_xi_t}
\end{center}
\end{figure} 

\begin{figure}
\begin{center}
\includegraphics[angle=0,width=0.4\textwidth]{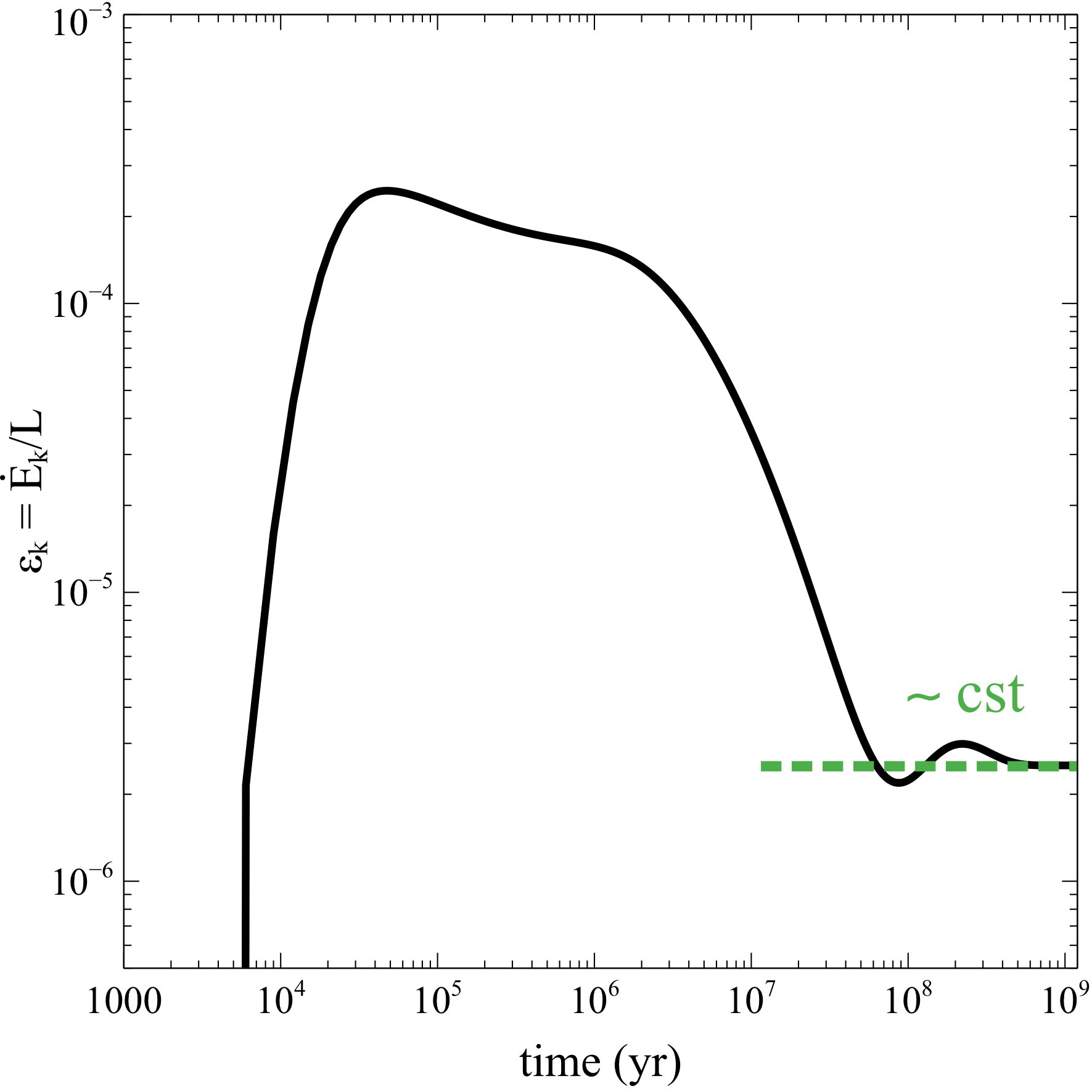} 
\caption{ Energy ratio versus time (black solid), with constant trend (green dotted). 
}
\label{Fig_K_t}
\end{center}
\end{figure}

In observational studies, three physical parameters are generally used to quantify the outflow energetics: the mass outflow rate ($\dot{M}$), the momentum flux ($\dot{p}$), and the kinetic power ($\dot{E}_k$). 
In the so-called thin shell approximation \citep[e.g.][]{Gonzalez-Alfonso_et_2017}, they are given by: 
\begin{equation}
\dot{M} = \frac{M_{g}}{t_{flow}} = M_{g} \frac{v}{r} = 4 \pi m_p R_0^2 n_0  v
\label{Eq_Mdot}
\end{equation} 
\begin{equation}
\dot{p}  = \dot{M} v= M_{g} \frac{v^2}{r} = 4 \pi m_p R_0^2 n_0  v^2
\label{Eq_pdot}
\end{equation} 
\begin{equation}
\dot{E}_k = \frac{1}{2} \dot{M} v^2 = \frac{1}{2} M_{g} \frac{v^3}{r} = 2 \pi m_p R_0^2 n_0  v^3
\label{Eq_Ekdot}
\end{equation}
where we introduce our assumption for the gas mass (Eq. \ref{Eq_gas_mass}) in the last equalities. 

In Fig. $\ref{Fig_MpEkdot_t}$, we show the corresponding temporal evolution of the outflow energetics. 
We see that all three outflow parameters reach a minimum around the Salpeter time ($t \gtrsim t_\mathrm{S}$), corresponding to the minimum in the shell velocity (cf. Fig. $\ref{Fig_v_t}$), and then start to grow exponentially. 
In fact, the outflow energetics scale with velocity as: $\dot{M} \propto v \propto \exp(t/3t_\mathrm{S})$, $\dot{p} \propto v^2 \propto \exp(2t/3t_\mathrm{S})$, $\dot{E}_k \propto v^3 \propto \exp(t/t_\mathrm{S})$. 
We note that the kinetic luminosity has the steepest dependence, equalling the luminosity output, whereas both the mass outflow rate and the momentum flux display a slower growth.  

Two derived quantities, the momentum ratio ($\zeta$) and the energy ratio ($\epsilon_k$), are also often used in quantifying the observed outflow energetics: 
\begin{equation}
\zeta = \frac{\dot{p}}{L/c} 
= \frac{4 \pi c m_p n_0 R_0^2 v^2}{L_0 \exp(t/t_S)}
\label{momentum_ratio}
\end{equation} 
\begin{equation}
\epsilon_k = \frac{\dot{E}_k}{L} 
= \frac{2 \pi m_p n_0 R_0^2 v^3}{L_0 \exp(t/t_S)}
\label{energy_ratio}
\end{equation}
where our assumptions are introduced in the last equalities. 

In Figs. $\ref{Fig_xi_t}$ and $\ref{Fig_K_t}$, we show the corresponding temporal evolution of the momentum ratio and energy ratio. We see that the momentum ratio falls off at late times, as $\zeta \propto \frac{exp(2t/3t_\mathrm{S})}{exp(t/t_\mathrm{S})} \propto exp(-t/3t_\mathrm{S})$. On the other hand, since the outflow kinetic power $\dot{E}_k(t)$ has the same dependence as the luminosity output $L(t)$ (both scaling as $\propto \exp(t/t_\mathrm{S})$), the energy ratio remains roughly constant in the exponential regime: $\epsilon_k \propto \frac{exp(t/t_\mathrm{S})}{exp(t/t_\mathrm{S})} \sim cst$. 


\subsection{Dependence on the underlying physical parameters and comparison with observations}

\begin{figure*}
\begin{multicols}{3}
    \includegraphics[width=\linewidth]{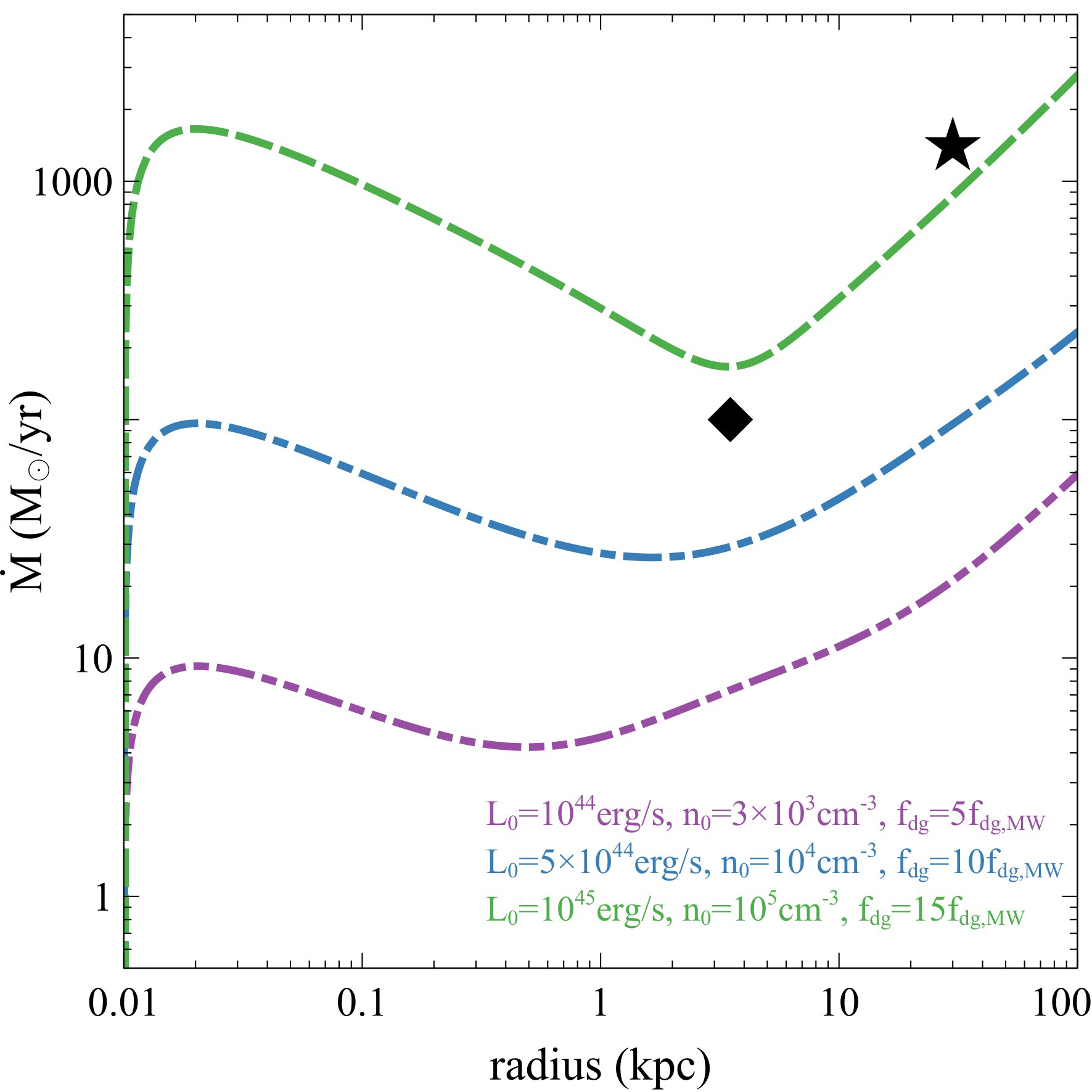}\par 
    \includegraphics[width=\linewidth]{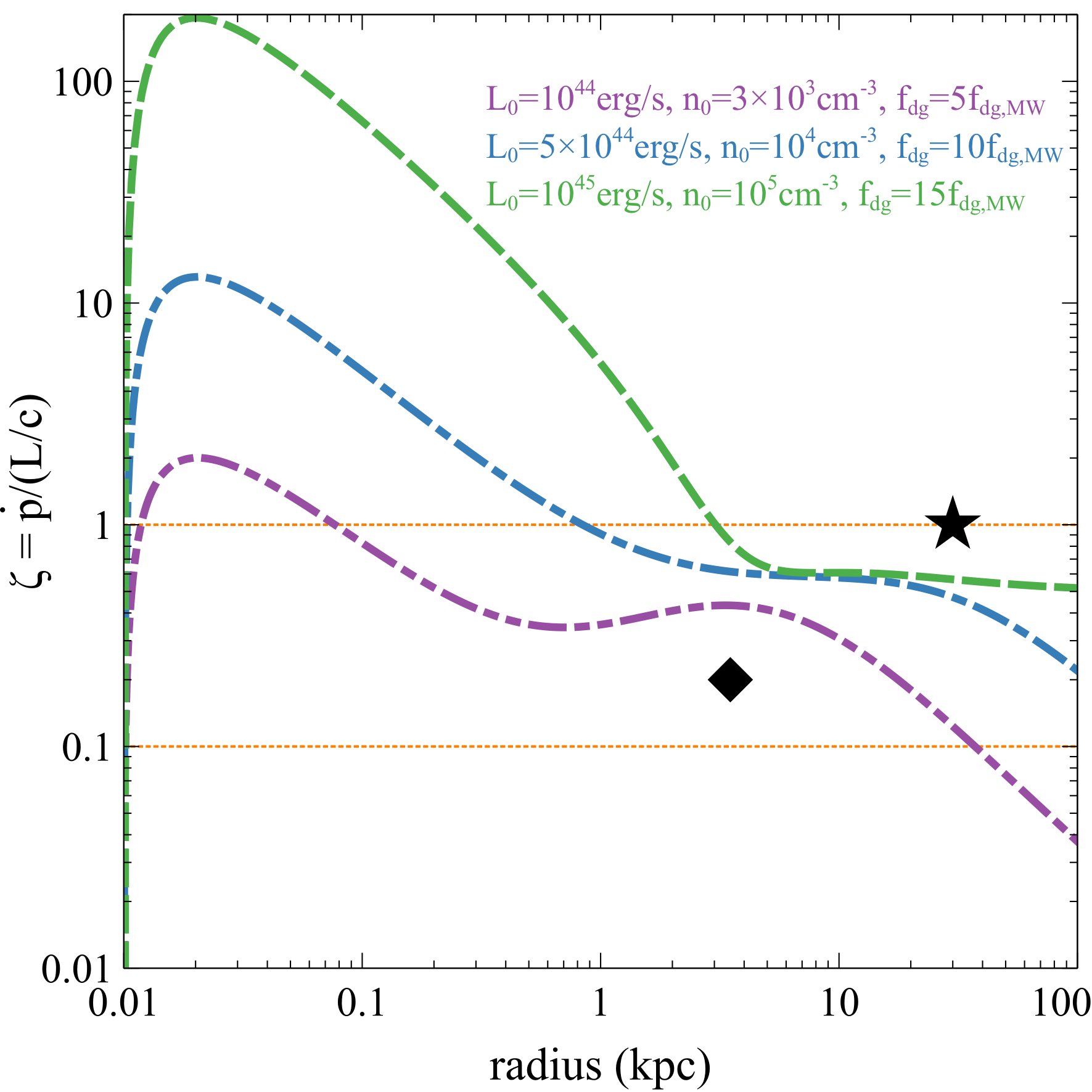}\par 
    \includegraphics[width=\linewidth]{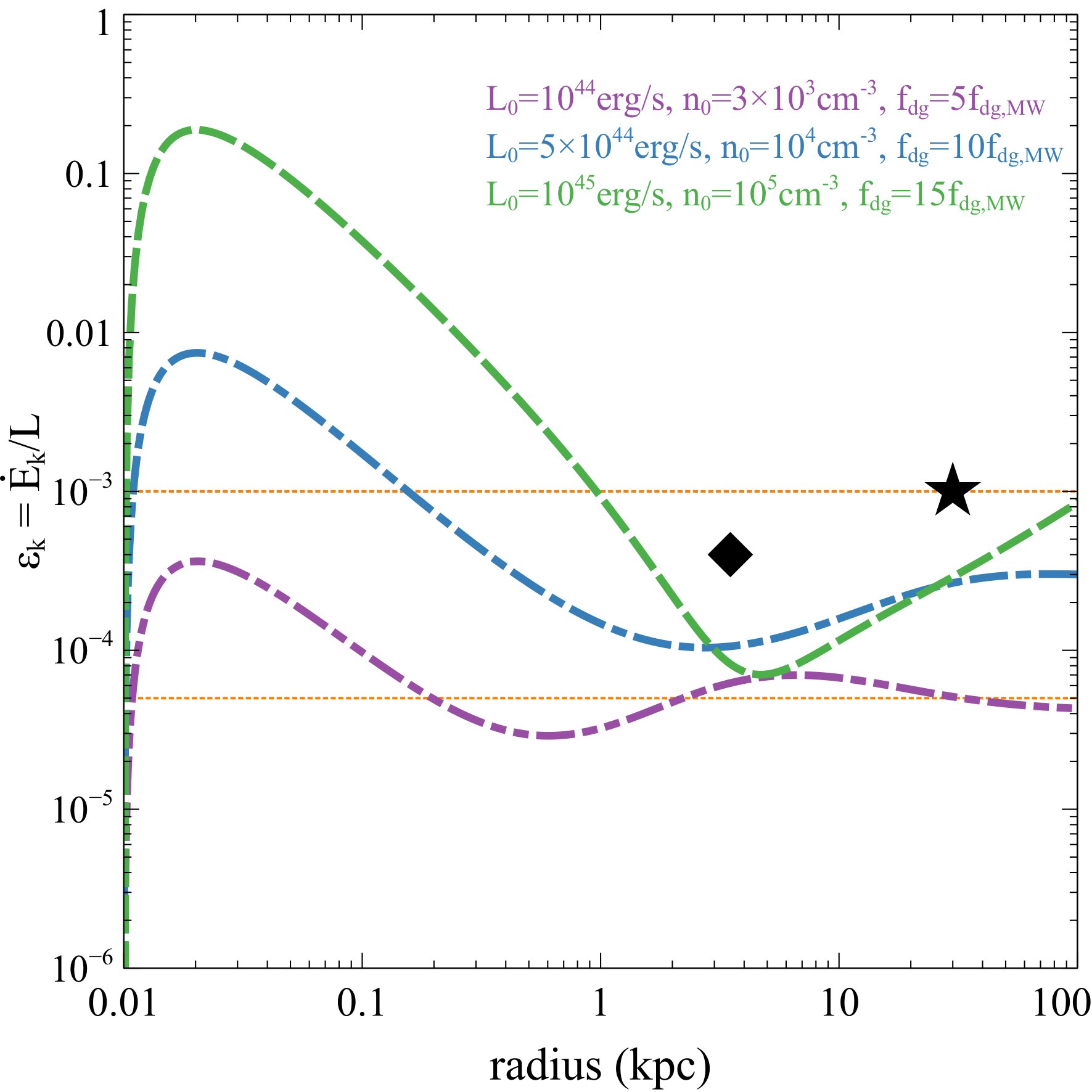}\par
    \end{multicols}
\caption{ 
Outflow energetics as a function of radius: mass outflow rate $\dot{M}$ (left panel), momentum ratio $\zeta$ (middle panel), and energy ratio $\epsilon_k$ (right panel). Variations in the model parameters: $L_0 = 10^{44}$ erg/s, $n_0 = 3 \times 10^3 cm^{-3}$, $f_{dg} = 5 \times f_\mathrm{dg,MW}$ (violet dash-dot-dot); $L_0 = 5 \times 10^{44}$ erg/s, $n_0 = 10^4 cm^{-3}$, $f_{dg} = 10 \times f_\mathrm{dg,MW}$ (blue dash-dot); $L_0 = 10^{45}$ erg/s, $n_0 = 10^5 cm^{-3}$, $f_{dg} = 15 \times f_\mathrm{dg,MW}$ (green dashed). 
Observational data: SDSS J1148+5251 (black star), the high-redshift QSO sample (black diamond), with typical ranges of momentum rate $(0.1-1) L/c$ and kinetic power $(0.005-0.1)\% L$ from \citet{Bischetti_et_2018} (orange fine dotted).
}
\label{Fig_energetics_t_model_obs}
\end{figure*}

In Fig. \ref{Fig_energetics_t_model_obs} we show the effects of varying the different physical parameters on the global outflow energetics. We observe that the momentum ratio is mostly lower than unity ($\zeta \lesssim 1$) on $\sim$kpc-scales, scaling as $\zeta \propto n_0 \kappa_{UV} r^{-1}$ in the exponential regime. 
Similarly, the energy ratio is mostly smaller than $\epsilon_k \lesssim 10^{-3}$ on galactic scales, scaling as $\epsilon_k \propto n_0 \kappa_{UV}$. We note that both $\zeta(r)$ and $\epsilon_k(r)$ directly scale with the external density and the dust-to-gas ratio: $\propto  n_0 f_{dg}$. This suggests that more powerful outflows can be driven in more extreme environments, characterised by high densities and large dust content, such as ULIRG-like sources \citep[cf.][]{Ishibashi_Fabian_2018}.

On the observational side, a number of quasar-driven outflows have been detected at very high redshifts.  
The prototype is the luminous quasar SDSS J1148+5251 at $z = 6.4$, in which a high-velocity outflow extending on large scales (up to $r \sim 30$ kpc) has been traced by the broad wings of the [CII] emission line \citep{Maiolino_et_2012, Cicone_et_2015}. The associated momentum rate and kinetic power are of the order of $\dot{p} \sim L/c$ and $\dot{E}_{k} \sim 10^{-3} L$. Most recently, quasar-driven outflows have been uncovered in a large sample of QSOs at $4.5 < z < 7.1$, through a stacking analysis of the broad but weak [CII] wings \citep{Bischetti_et_2018}. These outflows, with a typical spatial extent of $r \sim 3.5$ kpc, are characterised by momentum rates in the range $\sim(0.1-1) L/c$, and kinetic powers in the range $\sim(0.01-0.1)\%$ of $L$, respectively. For comparison, we plot in Fig. \ref{Fig_energetics_t_model_obs} the observational values of the outflow energetics for SDSS J1148+5251 (black star symbol) and the high-redshift QSO sample (black diamond symbol), as well as the typical range of momentum rates and kinetic powers reported in \citet{Bischetti_et_2018}.

\section{Discussion}

Recent observations confirm the existence of massive black holes ($M_{BH} \sim 10^9 M_{\odot}$), accreting at rates close to the Eddington limit, in the early Universe \citep{Banados_et_2018, Mazzucchelli_et_2017, Gallerani_et_2017}. The host galaxies of the luminous quasars at $z \sim 6-7$ contain considerable amounts of gas and dust, with dust masses in the range $M_{dust} \sim (10^7 - 10^9) M_{\odot}$ \citep{Venemans_et_2017, Venemans_et_2018}. 

The observed combination of high luminosity output and large dust content form favourable conditions for AGN radiative feedback, due to radiation pressure on dust. Here we explicitly analyse the evolution of AGN radiation pressure-driven outflows powered by Eddington-limited accretion with exponential luminosity output ($L \propto \exp(t/t_\mathrm{S})$). We find that such AGN radiative feedback can drive outflows on galactic scales, with the shell radius and velocity both following an exponential evolution of the form $\propto \exp(t/3t_\mathrm{S})$ at late times. The resulting outflow energetics are quite modest during the early growth phase, with typical values not exceeding $\zeta < 1$ and $\epsilon_k < 10^{-3}$. 

Observational measurements of the outflow energetics in the highest-redshift quasars yield typical values in the range: $\sim (0.1-1) L/c$ and $\sim (0.01-0.1)\%$ of $L$ \citep{Bischetti_et_2018}. 
Interestingly, the momentum rates and kinetic powers measured in the early quasars  tend to be significantly lower than the values reported for sources at lower redshifts.
In fact, the momentum fluxes and kinetic powers of local quasars typically span the range $\sim (1-30) L/c$ and $\sim (0.1-5)\%$ of $L$ \citep{Fiore_et_2017, Fluetsch_et_2019}. 
Moreover, the inferred energetics of early quasars are also much lower compared to predictions of wind-driven models, which suggest typical values of $\sim 20 L/c$ and $\sim 0.05 L$ in the wind energy-driven regime \citep{Zubovas_King_2012}. Considering the temporal evolution of the AGN luminosity output in the framework of the wind model, \citet{Zubovas_2018} shows that a power-law luminosity decay can adequately reproduce the observed range of momentum and energy loading factors, and at the same time preserve the correlations between outflow properties and AGN luminosity. Analysis of wind-driven outflows, powered by exponentially growing black holes, suggests that QSO-driven outflows are closer to the energy-driven limit than to the momentum-driven limit \citep{Gilli_et_2017}. It has also been argued that outflows, powered by tidal disruption events around massive black hole seeds, can noticeably affect the evolution of the host dwarf galaxies \citep{Zubovas_2019}. 

In the case of AGN radiative feedback, we naturally obtain low values of the momentum ratio and energy ratio during the early black hole growth phase, in agreement with observations of quasar-driven outflows at very high redshifts \citep{Bischetti_et_2018}. In fact, the modest outflow energetics observed in the early quasars may be easily accounted for in the framework of AGN radiative feedback. We note that our feedback model, based on radiation pressure on dust, is only effective at redshifts $z < z_{dust}$, once abundant dust production has occurred in the first galaxies. Early enrichment is suggested by the unexpectedly large amounts of dust observed in quasar host galaxies at $z > 6$, with a dust mass of $M_{dust} \gtrsim 10^8 M_{\odot}$ detected in the most distant quasar at $z = 7.5$ \citep{Venemans_et_2017}. It has been argued that the timescale for dust production in asymptotic giant branch (AGB) stars may be too slow compared to the young age of the Universe (e.g. $\sim 690$ Myr in the case of ULAS J1342+0928), and that other sources of dust are required at early times. In this context, dust production in supernova (SN) explosions has been often invoked, due to their shorter evolution timescales (typically a few $\sim$ Myr). Recent observations indicate that substantial quantities of dust can be released in core-collapse supernovae \citep{Owen_Barlow_2015, Wesson_et_2015}. Rapid dust formation by early supernovae ($z \gtrsim 7$) is required in order to account for the early dust enrichment observed near the reionization era \citep{Michalowski_2015, Watson_et_2015, Lesniewska_Michalowski_2019}. 

On the other hand, the presence of large amounts of dusty gas imply heavy obscuration in the first quasar host galaxies, suggesting that the early growth phase of massive black holes may be hidden from view. 
Indeed, the bulk of the AGN population is known to be obscured \citep{Fabian_1999}, and there could exist a whole population of highly accreting but heavily obscured quasars in the early Universe that currently elude detection \citep[][and references therein]{Hickox_Alexander_2018}. It is even possible that the outflows currently observed at very high $z$ \citep[e.g.][]{Bischetti_et_2018} only form the tip of the iceberg, and such quasar-driven outflows might be common in the early Universe. 
In our picture, AGN radiative feedback, which directly acts on the obscuring dusty gas, may help in clearing out the dusty gas from the early quasars, through the development of radiation pressure-driven outflows.  
However, the outflow energetics observed in the first quasars are significantly lower compared to typical values measured in local quasars. Taken at face value, the modest outflow energetics suggest that early radiative feedback may be less effective in clearing the entire host galaxy. In other words, the gas reservoir may not be completely disrupted, and the accretion process can potentially be resumed, allowing further black hole growth. 


\section*{Acknowledgements }

WI acknowledges support from the University of Zurich.

  
\bibliographystyle{mn2e}
\bibliography{biblio.bib}

\begin{thebibliography}{}

\bibitem[\protect\citeauthoryear{{Ba{\~n}ados}, {Venemans}, {Mazzucchelli},
  {Farina}, {Walter}, {Wang}, {Decarli}, {Stern}, {Fan}, {Davies}, {Hennawi},
  {Simcoe}, {Turner}, {Rix}, {Yang}, {Kelson}, {Rudie} \&
  {Winters}}{{Ba{\~n}ados} et~al.}{2018}]{Banados_et_2018}
{Ba{\~n}ados} E.,  {Venemans} B.~P.,  {Mazzucchelli} C.,  {Farina} E.~P.,
  {Walter} F.,  {Wang} F.,  {Decarli} R.,  {Stern} D.,  {Fan} X.,  {Davies}
  F.~B.,  {Hennawi} J.~F.,  {Simcoe} R.~A.,  {Turner} M.~L.,  {Rix} H.-W.,
  {Yang} J.,  {Kelson} D.~D.,  {Rudie} G.~C.,    {Winters} J.~M.,  2018, \nat,
  553, 473

\bibitem[\protect\citeauthoryear{{Bischetti}, {Maiolino}, {Fiore}, {Piconcelli}
  \& {Fluetsch}}{{Bischetti} et~al.}{2018}]{Bischetti_et_2018}
{Bischetti} M.,  {Maiolino} R.,  {Fiore} S.~C.~F.,  {Piconcelli} E.,
  {Fluetsch} A.,  2018, arXiv e-prints

\bibitem[\protect\citeauthoryear{{Calura}, {Gilli}, {Vignali}, {Pozzi},
  {Pipino} \& {Matteucci}}{{Calura} et~al.}{2014}]{Calura_et_2014}
{Calura} F.,  {Gilli} R.,  {Vignali} C.,  {Pozzi} F.,  {Pipino} A.,
  {Matteucci} F.,  2014, \mnras, 438, 2765

\bibitem[\protect\citeauthoryear{{Cicone}, {Maiolino}, {Gallerani}, {Neri},
  {Ferrara}, {Sturm}, {Fiore}, {Piconcelli} \& {Feruglio}}{{Cicone}
  et~al.}{2015}]{Cicone_et_2015}
{Cicone} C.,  {Maiolino} R.,  {Gallerani} S.,  {Neri} R.,  {Ferrara} A.,
  {Sturm} E.,  {Fiore} F.,  {Piconcelli} E.,    {Feruglio} C.,  2015, \aap,
  574, A14

\bibitem[\protect\citeauthoryear{{Fabian}}{{Fabian}}{1999}]{Fabian_1999}
{Fabian} A.~C.,  1999, \mnras, 308, L39

\bibitem[\protect\citeauthoryear{{Fiore}, {Feruglio}, {Shankar}, {Bischetti},
  {Bongiorno}, {Brusa}, {Carniani}, {Cicone}, {Duras}, {Lamastra}, {Mainieri},
  {Marconi}, {Menci}, {Maiolino}, {Piconcelli}, {Vietri} \&
  {Zappacosta}}{{Fiore} et~al.}{2017}]{Fiore_et_2017}
{Fiore} F.,  {Feruglio} C.,  {Shankar} F.,  {Bischetti} M.,  {Bongiorno} A.,
  {Brusa} M.,  {Carniani} S.,  {Cicone} C.,  {Duras} F.,  {Lamastra} A.,
  {Mainieri} V.,  {Marconi} A.,  {Menci} N.,  {Maiolino} R.,  {Piconcelli} E.,
  {Vietri} G.,    {Zappacosta} L.,  2017, \aap, 601, A143

\bibitem[\protect\citeauthoryear{{Fluetsch}, {Maiolino}, {Carniani}, {Marconi},
  {Cicone}, {Bourne}, {Costa}, {Fabian}, {Ishibashi} \& {Venturi}}{{Fluetsch}
  et~al.}{2019}]{Fluetsch_et_2019}
{Fluetsch} A.,  {Maiolino} R.,  {Carniani} S.,  {Marconi} A.,  {Cicone} C.,
  {Bourne} M.~A.,  {Costa} T.,  {Fabian} A.~C.,  {Ishibashi} W.,    {Venturi}
  G.,  2019, \mnras, 483, 4586

\bibitem[\protect\citeauthoryear{{Gallerani}, {Fan}, {Maiolino} \&
  {Pacucci}}{{Gallerani} et~al.}{2017}]{Gallerani_et_2017}
{Gallerani} S.,  {Fan} X.,  {Maiolino} R.,    {Pacucci} F.,  2017, \pasa, 34,
  e022

\bibitem[\protect\citeauthoryear{{Gilli}, {Calura}, {D'Ercole} \&
  {Norman}}{{Gilli} et~al.}{2017}]{Gilli_et_2017}
{Gilli} R.,  {Calura} F.,  {D'Ercole} A.,    {Norman} C.,  2017, \aap, 603, A69

\bibitem[\protect\citeauthoryear{{Gonz{\'a}lez-Alfonso}, {Fischer}, {Spoon},
  {Stewart}, {Ashby}, {Veilleux}, {Smith}, {Sturm} \& et
  al.}{{Gonz{\'a}lez-Alfonso} et~al.}{2017}]{Gonzalez-Alfonso_et_2017}
{Gonz{\'a}lez-Alfonso} E.,  {Fischer} J.,  {Spoon} H.~W.~W.,  {Stewart} K.~P.,
  {Ashby} M.~L.~N.,  {Veilleux} S.,  {Smith} H.~A.,  {Sturm} E.,    et al.
  2017, \apj, 836, 11

\bibitem[\protect\citeauthoryear{{Hickox} \& {Alexander}}{{Hickox} \&
  {Alexander}}{2018}]{Hickox_Alexander_2018}
{Hickox} R.~C.,  {Alexander} D.~M.,  2018, \araa, 56, 625

\bibitem[\protect\citeauthoryear{{Ishibashi} \& {Fabian}}{{Ishibashi} \&
  {Fabian}}{2015}]{Ishibashi_Fabian_2015}
{Ishibashi} W.,  {Fabian} A.~C.,  2015, \mnras, 451, 93

\bibitem[\protect\citeauthoryear{{Ishibashi} \& {Fabian}}{{Ishibashi} \&
  {Fabian}}{2018}]{Ishibashi_Fabian_2018}
{Ishibashi} W.,  {Fabian} A.~C.,  2018, \mnras, 481, 4522

\bibitem[\protect\citeauthoryear{{Ishibashi}, {Fabian} \&
  {Maiolino}}{{Ishibashi} et~al.}{2018}]{Ishibashi_et_2018}
{Ishibashi} W.,  {Fabian} A.~C.,    {Maiolino} R.,  2018, \mnras, 476, 512

\bibitem[\protect\citeauthoryear{{Le{\'s}niewska} \&
  {Micha{\l}owski}}{{Le{\'s}niewska} \&
  {Micha{\l}owski}}{2019}]{Lesniewska_Michalowski_2019}
{Le{\'s}niewska} A.,  {Micha{\l}owski} M.~J.,  2019, \aap, 624, L13

\bibitem[\protect\citeauthoryear{{Maiolino}, {Gallerani}, {Neri}, {Cicone},
  {Ferrara}, {Genzel}, {Lutz}, {Sturm}, {Tacconi}, {Walter}, {Feruglio},
  {Fiore} \& {Piconcelli}}{{Maiolino} et~al.}{2012}]{Maiolino_et_2012}
{Maiolino} R.,  {Gallerani} S.,  {Neri} R.,  {Cicone} C.,  {Ferrara} A.,
  {Genzel} R.,  {Lutz} D.,  {Sturm} E.,  {Tacconi} L.~J.,  {Walter} F.,
  {Feruglio} C.,  {Fiore} F.,    {Piconcelli} E.,  2012, \mnras, 425, L66

\bibitem[\protect\citeauthoryear{{Mazzucchelli}, {Ba{\~n}ados}, {Venemans},
  {Decarli}, {Farina}, {Walter}, {Eilers} \& et al.}{{Mazzucchelli}
  et~al.}{2017}]{Mazzucchelli_et_2017}
{Mazzucchelli} C.,  {Ba{\~n}ados} E.,  {Venemans} B.~P.,  {Decarli} R.,
  {Farina} E.~P.,  {Walter} F.,  {Eilers} A.-C.,    et al. 2017, \apj, 849, 91

\bibitem[\protect\citeauthoryear{{Micha{\l}owski}}{{Micha{\l}owski}}{2015}]{Michalowski_2015}
{Micha{\l}owski} M.~J.,  2015, \aap, 577, A80

\bibitem[\protect\citeauthoryear{{Mortlock}, {Warren}, {Venemans}, {Patel},
  {Hewett}, {McMahon}, {Simpson}, {Theuns}, {Gonz{\'a}les-Solares}, {Adamson},
  {Dye}, {Hambly}, {Hirst}, {Irwin}, {Kuiper}, {Lawrence} \&
  {R{\"o}ttgering}}{{Mortlock} et~al.}{2011}]{Mortlock_et_2011}
{Mortlock} D.~J.,  {Warren} S.~J.,  {Venemans} B.~P.,  {Patel} M.,  {Hewett}
  P.~C.,  {McMahon} R.~G.,  {Simpson} C.,  {Theuns} T.,  {Gonz{\'a}les-Solares}
  E.~A.,  {Adamson} A.,  {Dye} S.,  {Hambly} N.~C.,  {Hirst} P.,  {Irwin}
  M.~J.,  {Kuiper} E.,  {Lawrence} A.,    {R{\"o}ttgering} H.~J.~A.,  2011,
  \nat, 474, 616

\bibitem[\protect\citeauthoryear{{Owen} \& {Barlow}}{{Owen} \&
  {Barlow}}{2015}]{Owen_Barlow_2015}
{Owen} P.~J.,  {Barlow} M.~J.,  2015, ArXiv e-prints

\bibitem[\protect\citeauthoryear{{Smith}, {Bromm} \& {Loeb}}{{Smith}
  et~al.}{2017}]{Smith_et_2017}
{Smith} A.,  {Bromm} V.,    {Loeb} A.,  2017, Astronomy and Geophysics, 58,
  3.22

\bibitem[\protect\citeauthoryear{{Valiante}, {Agarwal}, {Habouzit} \&
  {Pezzulli}}{{Valiante} et~al.}{2017}]{Valiante_et_2017}
{Valiante} R.,  {Agarwal} B.,  {Habouzit} M.,    {Pezzulli} E.,  2017, \pasa,
  34, e031

\bibitem[\protect\citeauthoryear{{Venemans}, {Decarli}, {Walter},
  {Ba{\~n}ados}, {Bertoldi}, {Fan}, {Farina}, {Mazzucchelli}, {Riechers},
  {Rix}, {Wang} \& {Yang}}{{Venemans} et~al.}{2018}]{Venemans_et_2018}
{Venemans} B.~P.,  {Decarli} R.,  {Walter} F.,  {Ba{\~n}ados} E.,  {Bertoldi}
  F.,  {Fan} X.,  {Farina} E.~P.,  {Mazzucchelli} C.,  {Riechers} D.,  {Rix}
  H.-W.,  {Wang} R.,    {Yang} Y.,  2018, \apj, 866, 159

\bibitem[\protect\citeauthoryear{{Venemans}, {Walter}, {Decarli},
  {Ba{\~n}ados}, {Carilli}, {Winters}, {Schuster}, {da Cunha}, {Fan}, {Farina},
  {Mazzucchelli}, {Rix} \& {Weiss}}{{Venemans} et~al.}{2017}]{Venemans_et_2017}
{Venemans} B.~P.,  {Walter} F.,  {Decarli} R.,  {Ba{\~n}ados} E.,  {Carilli}
  C.,  {Winters} J.~M.,  {Schuster} K.,  {da Cunha} E.,  {Fan} X.,  {Farina}
  E.~P.,  {Mazzucchelli} C.,  {Rix} H.-W.,    {Weiss} A.,  2017, \apjl, 851, L8

\bibitem[\protect\citeauthoryear{{Volonteri}}{{Volonteri}}{2010}]{Volonteri_2010}
{Volonteri} M.,  2010, \aapr, 18, 279

\bibitem[\protect\citeauthoryear{{Wang}, {Yang}, {Fan}, {Wu}, {Yue}, {Li},
  {Bian}, {Jiang} \& {Ba{\~n}ados}}{{Wang} et~al.}{2018}]{Wang_et_2018}
{Wang} F.,  {Yang} J.,  {Fan} X.,  {Wu} X.-B.,  {Yue} M.,  {Li} J.-T.,  {Bian}
  F.,  {Jiang} L.,    {Ba{\~n}ados} e.~a.,  2018, arXiv e-prints

\bibitem[\protect\citeauthoryear{{Watson}, {Christensen}, {Knudsen}, {Richard},
  {Gallazzi} \& {Micha{\l}owski}}{{Watson} et~al.}{2015}]{Watson_et_2015}
{Watson} D.,  {Christensen} L.,  {Knudsen} K.~K.,  {Richard} J.,  {Gallazzi}
  A.,    {Micha{\l}owski} M.~J.,  2015, \nat, 519, 327

\bibitem[\protect\citeauthoryear{{Wesson}, {Barlow}, {Matsuura} \&
  {Ercolano}}{{Wesson} et~al.}{2015}]{Wesson_et_2015}
{Wesson} R.,  {Barlow} M.~J.,  {Matsuura} M.,    {Ercolano} B.,  2015, \mnras,
  446, 2089

\bibitem[\protect\citeauthoryear{{Wise}}{{Wise}}{2018}]{Wise_2018}
{Wise} J.~H.,  2018, arXiv e-prints

\bibitem[\protect\citeauthoryear{{Wu}, {Wang}, {Fan}, {Yi}, {Zuo}, {Bian},
  {Jiang}, {McGreer}, {Wang}, {Yang}, {Yang}, {Thompson} \& {Beletsky}}{{Wu}
  et~al.}{2015}]{Wu_et_2015}
{Wu} X.-B.,  {Wang} F.,  {Fan} X.,  {Yi} W.,  {Zuo} W.,  {Bian} F.,  {Jiang}
  L.,  {McGreer} I.~D.,  {Wang} R.,  {Yang} J.,  {Yang} Q.,  {Thompson} D.,
  {Beletsky} Y.,  2015, \nat, 518, 512

\bibitem[\protect\citeauthoryear{{Yang}, {Wang}, {Fan}, {Yue}, {Wu}, {Li},
  {Bian}, {Jiang}, {Ba{\~n}ados} \& {Beletsky}}{{Yang}
  et~al.}{2018}]{Yang_et_2018}
{Yang} J.,  {Wang} F.,  {Fan} X.,  {Yue} M.,  {Wu} X.-B.,  {Li} J.,  {Bian} F.,
   {Jiang} L.,  {Ba{\~n}ados} E.,    {Beletsky} Y.,  2018, arXiv e-prints

\bibitem[\protect\citeauthoryear{{Zubovas}}{{Zubovas}}{2018}]{Zubovas_2018}
{Zubovas} K.,  2018, \mnras, 473, 3525

\bibitem[\protect\citeauthoryear{{Zubovas}}{{Zubovas}}{2019}]{Zubovas_2019}
{Zubovas} K.,  2019, \mnras, 483, 1957

\bibitem[\protect\citeauthoryear{{Zubovas} \& {King}}{{Zubovas} \&
  {King}}{2012}]{Zubovas_King_2012}
{Zubovas} K.,  {King} A.,  2012, \apjl, 745, L34

\end{thebibliography}


\label{lastpage}

\end{document}